\documentclass[aps,prb,twocolumn,showpacs,superscriptaddress]{revtex4-1}
\usepackage{graphicx}
\usepackage{dcolumn}
\usepackage{bm}
\usepackage{amsmath}
\begin{document}
\title{First-principles study of the lattice instabilities in
Mn$_{2}$Ni{\it X} ({\it X}= Al, Ga, In, Sn) magnetic shape memory
alloys}
\author{Souvik Paul}%
\affiliation{Department of Physics, Indian Institute of Technology
Guwahati, Guwahati, Assam 781039, India}%
\author{Biplab Sanyal}
\affiliation{Department of Physics and Astronomy,
Uppsala University, Box 516, 75120 Uppsala, Sweden}
\author{ Subhradip Ghosh}%
\email{For correspondence: subhra@iitg.ernet.in}%
\affiliation{Department of Physics, Indian Institute of Technology
Guwahati, Guwahati, Assam 781039, India}%

\date{\today}

\begin{abstract}
Using first-principles based Density Functional Theory (DFT),
we have investigated the structural instabilities in the
austenite phases of Mn$_{2}$Ni{\it X} ({\it X}= Al, Ga, In, Sn)
magnetic shape memory alloys (MSMA).
A complete softening is observed in the acoustic TA$_{2}$ branches
for all the materials along [$\xi\xi$0] directions leading to
the instability in the austenite structure which effectively
stabilizes into martensitic structure. The reasons behind this
softening are traced back to the repulsion from the optical T$_{2g}$
branches and to the nesting features in the Fermi surfaces. The vibrational density
of states, the force constants and the elastic moduli are also computed
and analyzed, which reconfirm the underlying mechanism behind the instabilities.
The results indicate that the phonon anomalies are related to the occurrence
of possible pre-martensitic phases which can be quite complex.
\end{abstract}

\maketitle
\section{Introduction}
Magnetic shape memory alloys (MSMA) are excellent candidates for
technological applications due to their coupling between different
degrees of freedom, such as caloric, magnetic, elastic etc., that
introduce multifunctionality in these materials. The magneto-structural
coupling results in a phase transition between high temperature
austenite structure and low temperature martensitic variants driven
by magnetic field under ambient conditions. Microscopically, this
martensitic transformation is merely a consequence of reshuffling of atomic planes,
which is often mediated through different periodically modulated meta-stable
structures called premartensitic structure.

Very often, the microscopic origin  behind the martensitic phase transformation can be
explained by softening of some phonon modes, related to the softness in
elastic stiffness constants caused by the  nesting topology between
parallel Fermi surfaces due to intense electron-phonon coupling.This has been
the case for almost all the ternary MSMA's crystallizing in Heusler structure.
The most
extensively studied ternary MSMA is Ni$_{2}$MnGa, which in single crystal environments
and close to the stoichiometric composition, exhibit nearly 10$\%$ magnetic
field-induced strains (MFIS) under a magnetic
field of less than 1 Tesla\cite{nimnga1,nimnga2}, making it a strong contender for
micro-mechanical sensors and actuators.
The structural instability of Ni$_{2}$MnGa in the austenite phase has been
linked with an anomalous phonon softening of transverse acoustic
TA$_{2}$ branch along [$\xi\xi0$] direction. The softening occurs at a
fractional wave vector $\xi$=(0.33,0.33,0) and it becomes more
prominent as one approaches towards the martensitic phase with decreasing
temperature\cite{ps1,ps2,ps3,ps4}.
The phonon softening has been found to correlate with the
premartensitic phase, which is anticipated by the precursor phonon softening
at that wave vector. Inelastic neutron scattering experiments and elastic
constants measurements on the high temperature phase corroborated the theoretical
calculations of softening of acoustic branch. A complete softening of
acoustic TA$_{2}$ branch along [$\xi\xi0$] direction with unstable phonon modes
was reported theoretically in Ni$_{2}$MnAl\cite{nimnaltheo}.
Later, Moya {\it et al.} verified the Kohn anomaly observed
in the theoretical study of TA$_{2}$ branch in this material by performing inelastic neutron scattering
experiment on nearly stoichiometric Ni$_{2}$MnAl \cite{nimnalexp}.
However, the experimental
softening is not complete since the phonon frequencies remain finite even
at lowest temperature. This could be
related to the fact that the composition needed for the martensitic phase
transformation to occur in Ni$_{2}$MnAl is slightly
off-stoichiometric\cite{nimnaloffsti}.
First-principles calculations observed similar phonon softening
of the same acoustic branch in Ni$_{2}$MnIn, Ni$_{2}$MnSb, and
Ni$_{2}$MnSn\cite{nimnin,nimnsn-sb}.

Although Ni$_{2}$MnGa near the stoichiometric composition is the first discovered
ternary system in Heusler structure exhibiting magnetic shape memory effect, and has
been studied extensively revealing a lot of interesting physics, its use in practical
applications is hindered due to the  martensitic transformation temperature being lower
than the room temperature, and poor ductility in poly-crystalline phase \cite{web,nimnga3}.
Attempts were made to improve
the functionality of the material by introducing disorder with various
possibilities and replacing Ga conjointly with Al, Ge, In, Sn and Sb.
However, the yield is not as fruitful as expected. Therefore, a quest for new
MSMA began with higher operating temperatures and better elastic properties
compared to Ni$_{2}$Mn{\it X}.
Recently, Mn$_{2}$NiGa has been reported to be a MSMA with promising
functional properties \cite{mn2niga,mn2niga1,mn2niga2,mn2niga3}. It has a martensitic
transformation temperature close to room temperature (270 K) and much
broader hysteresis loop \cite{mn2niga}. An excellent two-way shape memory effect with strains
of 1.7$\%$ and field controllable shape memory effect up to 4$\%$ has been
observed experimentally in single crystalline environment \cite{mn2niga}.
Experiments with poly-crystalline sample found the martensitic transformation
temperature to be at 230 K \cite{mn2niga2}. It also observed that the structural
transformation is dependent upon residual stress. According to the analysis of
their Powder X-ray diffraction data, the system undergoes a martensitic
transformation to either a non-modulated tetragonal structure or a monoclinic
modulated structure at room temperature, depending upon the residual stress.
Neutron Powder diffraction experiments on this system confirmed the presence
of an orthorhombic modulated structure, which is independent of temperature \cite{mn2niga3}.
These results generated interests in this system in the context of understanding its
structural stability and connections to shape memory effect. Another reason is that this
material draws attention due to its relatively
high Curie temperature (T$_{C}\sim$ 588 K) \cite{mn2niga}
compared to Ni$_{2}$MnGa. In this respect, it is notable that Mn$_{2}$NiSn also
has a high T$_{C}$ (530 K)\cite{tcmn2nisn}. Driven by the possibilities of realizing
new MSMAs with functionalities better than the Ni$_{2}$MnX ones, first-principles
electronic structure calculations have been done on
Mn$_{2}$Ni{\it X} ({\it X}= Al, Ga, In, Sn) systems
\cite{pauljap,mn2nigatheo,mn2nialtheo,mn2niintheo,paulmn2nisn}. The results are quite encouraging
as the total energy calculations predicted  transformations from cubic austenite to
a non-modulated tetragonal phase at low temperatures conserving the volume, a signature
of shape-memory property. These results thus open up the possibility to further
investigate the origin of such transformations and their consequences in these
materials.

In this paper, we, therefore, make an attempt to understand the physical origin behind
the transformations by examining the vibrational properties of these materials in a
systematic way. We compute the phonon dispersion, the vibrational density of states,
the elastic constants and the Fermi surfaces in order to see whether connections to
the martensitic transformations can be made for these materials. The paper is organized
as follows: in section II, we provide details of the computational methods used, in section
III, we discuss the phonon dispersion relations, the vibrational densities of states,
the inter-atomic force constants, the elastic constants and the Fermi surfaces in
order to ascertain the mechanisms driving the martensitic transformations and finally
we summarize our results indicating their relevance for future research.

\section{Computational Details}
The electronic structures of the systems considered
were calculated using the Plane-Wave Pseudopotential (PW-PP) formalism
of the Density Functional Theory (DFT),
as implemented in \textsc{Quantum Espresso}\cite{qe} .
UltraSoft Pseudo Potentials (USPP)\cite{uspp}
were used to accurately calculate the electronic ground
states. Spin polarized Generalized
Gradient Approximation (GGA) scheme was used as the exchange-correlation
part of the potential with Perdew-Wang 91 parameterizations (PW91)\cite{pw91}.
Plane waves with energies up to 544 eV were used to describe
electronic wave functions. Fourier component of the augmented
charge density with cut-off energy up to 6530 eV was taken after
convergence tests. The Brillouin zone integrations were
carried out with finite temperature Methfessel-Paxton smearing\cite{mp}
method using 12$\times$12$\times$12 uniform $\it {k}$-mesh, which effectively
leads to 364 $\it {k}$-points in the irreducible wedge of the Brillouin zone.
The value of the smearing parameter was taken as 0.27 eV. Such choices
of the parameters ensure the convergence of phonon frequencies within 5$\%$.

The phonon dispersion relations were computed using Density Functional
Perturbation Theory (DFPT)\cite{dfpt}.
The DFPT scheme
is employed to accurately calculate the dynamical properties in condensed
matter systems with the precision at par with the electronic structure calculations.
The energy threshold value for convergence was 10$^{-16}$ Ry in phonon calculations.
Dynamical matrices were conveniently calculated in reciprocal
space from the ground state charge density and from its linear response to
the distortion in the ionic configurations. Fourier transform was employed thereafter
to obtain the real space force constants. The dynamical matrices were calculated
in a 4$\times$4$\times$4 $\it {q}$-point grid for all the structures. Convergence
of phonon frequencies within $1-2 \%$ was ensured by comparing frequencies calculated directly
and frequencies obtained by the Fourier transform of the dynamical matrices. Such
convergence tests ensured accuracy in elastic constants as they are calculated from
the slopes of the phonon dispersion curves.
The Fermi surfaces were calculated on $24\times24\times24$ highly
dense uniform $\it {k}$-point grid. It may be noted that the strength of the phonon anomaly is extremely
sensitive to temperature. An increase in temperature can reduce the nesting
features of Fermi surfaces and thus weaken the anomaly. In
DFT based calculations, the smearing parameter $\sigma$ plays the role of
fictitious electronic temperature. Therefore, to reduce the effect of finite
temperature in the calculations of Fermi surfaces, we kept $\sigma$=0.01 eV all along.

\section{Results and Discussions}
\subsection{Phonon dispersion}
\begin{figure}[h]
\includegraphics[width=8.5cm]{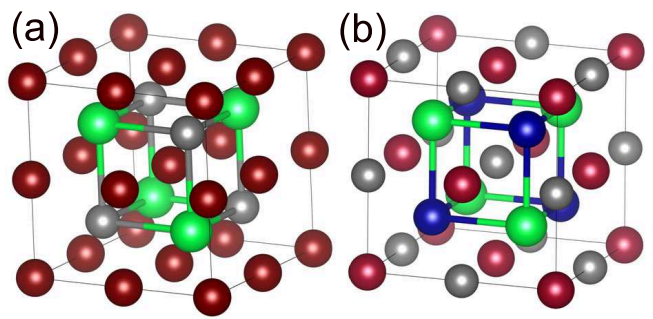}
\caption{(Color online) (a) The fcc L2$_{1}$ usual Heusler structure of
Mn$_{2}$Ni{\it X} (Ni$_{2}$Mn{\it X}) systems. The red, gray and green spheres
represents Mn (Ni), Ni (Mn) and {\it X} ({\it X}) atoms, respectively. (b) The
fcc Hg$_{2}$CuTi inverse Heusler structure of Mn$_{2}$Ni{\it X} systems. The red,
blue, gray and green spheres represent MnI, MnII, Ni and {\it X} atoms,
respectively.}
\end{figure}

\begin{figure*}
\includegraphics[width=5in]{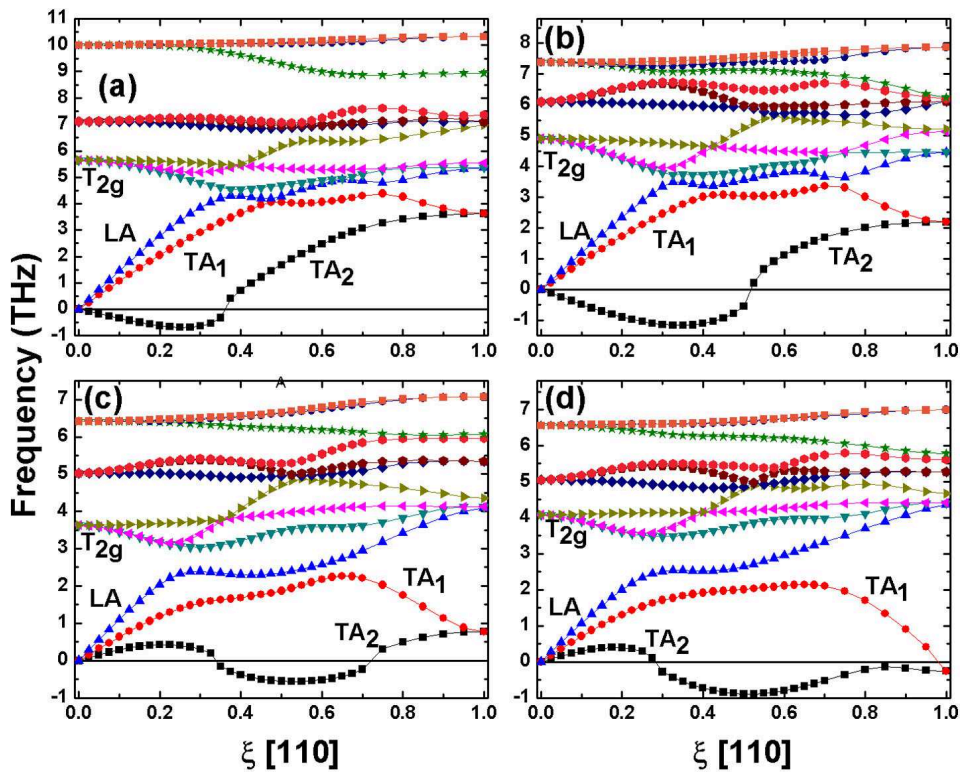}
\caption{(Color online) Phonon dispersion relations of
(a) Mn$_{2}$NiAl, (b) Mn$_{2}$NiGa, (c) Mn$_{2}$NiIn and
(d) Mn$_{2}$NiSn along [$\xi\xi$0] highly symmetric direction
of the Brillouin zone. The phonon wave vector $\xi$ is expressed
in units of ($\frac{2\pi}{a}$).}
\end{figure*}
Experimental measurements \cite{mn2niga,mn2niga1,mn2niga2,mn2nisnexp}and theoretical calculations
\cite{pauljap,mn2nigatheo,mn2niintheo,mn2nialtheo}have confirmed
that the alloys considered here favor Hg$_{2}$CuTi structure (Space group $F\bar{4}3m$), also
known as inverse Heusler structure, in the cubic austenite phase as opposed to
the usual Heusler structure of Ni$_{2}$Mn{\it X}.
The latter structure is best visualized as
four interpenetrating f.c.c sub-lattices at (0,0,0), (0.25,0.25,0.25),
(0.50,0.50,0.50) and (0.75,0.75,0.75), where the first and the third
positions are occupied by Mn atoms, second and the fourth positions by
Ni and {\it X} atoms, respectively(FIG. 1(a)). Interchanging the tetrahedral Mn
atom at (0.50,0.50,0.50) with octahedral Ni atom at (0.25,0.25,0.25)
keeping the remaining atoms fixed at their positions, leads to inverse
Heusler structure (FIG. 1(b)). Hereafter, Mn atom at (000) sub-lattice will be
denoted as MnI and the one at (0.25,0.25,0.25) as MnII.

Due to the
unavailability of experimental results on the lattice constants of Mn$_{2}$NiAl and
Mn$_{2}$NiIn, we have calculated the
equilibrium lattice constants of all the four materials with GGA
exchange-correlation functional and used them here. The total energies as a function
of lattice parameters were fitted to Murnaghan equation of state to
accurately calculate the equilibrium lattice constants. Our calculated
lattice constant for Mn$_{2}$NiGa is 5.85 \AA\ and for Mn$_{2}$NiSn is
6.15 \AA\, which agree well with the available experimental results,
i.e., 5.90 \AA\ for Mn$_{2}$NiGa \cite{mn2niga1}
and 6.1 \AA\ for Mn$_{2}$NiSn\cite{mn2nisnexp}.
On the other hand, our calculated lattice constants for Mn$_{2}$NiIn (a= 6.16 \AA)
matches well with available theoretical result\cite{mn2niintheo}.
Since the experimental results,
for Mn$_{2}$NiGa and Mn$_{2}$NiSn,
agree well with our calculated results, we consider our lattice
constants as good representations of the experimental ones.
The phonon dispersion spectra calculated
at those lattice constants along [$\xi\xi0$] highly symmetric direction
in the irreducible segment of the Brillouin zone (IBZ) are shown in FIG. 2.
The main interest lies in the transverse acoustic TA$_{2}$ branch, which
exists due to the atomic displacements [$\xi\bar{\xi}$0]
perpendicular to the propagation direction [$\xi\xi$0].
For all Heusler systems exhibiting martensitic transformation,
this branch shows an anomalous behavior. Therefore,
our aim is to investigate the behavior of acoustic TA$_{2}$ branch
along [$\xi\xi0$] direction.
The most important features in the dispersion curves are the anomalous dips
of the acoustic TA$_{2}$ branches where the phonon frequencies become imaginary,
suggesting instabilities in the cubic austenite structures which
usher in a phase transition to stable martensitic phases in all four materials.
In Mn$_{2}$NiGa and Mn$_{2}$NiAl,
the acoustic TA$_{2}$ branches have negative slopes at $\Gamma$ point,
indicating a pure elastic instability in their parent structure.
The range of this instability extends up to $\xi$=0.50 for Mn$_{2}$NiGa
and up to $\xi$=0.35 for Mn$_{2}$NiAl. The maximum of the dip  occur at
wave vectors $\xi$=0.35 and $\xi$=0.25 for Mn$_{2}$NiGa and
Mn$_{2}$NiAl, respectively. For Mn$_{2}$NiIn,
the instability of TA$_{2}$ branch starts from $\xi$=0.3
producing
maximum of the dip at wave vector $\xi$=0.50. For Mn$_{2}$NiSn,
unlike the other materials, the softening extends up to the wedge of the
Brillouin zone with the maximum of the dip at $\xi$=0.50.

In previous studies of lattice dynamics on ternary MSMAs with Heusler structures,
phonon anomalies
of TA$_{2}$ were correlated with the precursor phenomenon prior to the martensitic
phase when the systems are cooled from high temperatures. The wave vectors
corresponding to the imaginary phonon frequencies indicated shuffling
of atomic planes which stabilize the (c/a)$<$1 phases compared to the parent
phase ((c/a)$=$1). The occurrence of 3M, 5M and 7M modulated structures and even
incommensurate structures were confirmed experimentally. Possibilities of such modulated
structures can be inferred from the anomalies in our calculated dispersion relations
for Mn$_{2}$Ni{\it X} systems. A modulated structure with a periodicity of 8 atomic planes
(2M structure) can be associated with an instability at $\xi$=0.25,
one with a periodicity of 6 atomic
planes (3M structure) can be associated with an instability at $\xi$=0.33 and one
with a shuffling of 14 atomic planes (7M structure) can be associated with an instability
at $\xi$=0.29.
For Mn$_{2}$NiAl, the unstable mode occurs for $\xi$=0.0 to $\xi$=0.35 with
the maximum of the dip
at $\xi$=0.25 . This suggests the possibilities of occurrence of several modulated
phases.
The commensurate wave vector closest to the maximum of the dip in the TA$_{2}$
branch of of Mn$_{2}$NiGa
occurs at $\xi$=0.33 which can be related to the occurrence of the 3M structure.
Since, in Mn$_{2}$NiGa, the imaginary frequencies extends up to $\xi$=0.50, in
addition to aforementioned modulated structures 5M, modulation can also be observed
at $\xi$=0.43 which stabilizes with the shuffling of 10 atomic planes. In cases
of Mn$_{2}$NiIn and Mn$_{2}$NiSn, the maximum in the dip of the TA$_{2}$ branch occurs
at $\xi$=0.5, which although cannot be connected to the known modulated
structures mentioned above, but the extent of the instabilities in these systems
can be connected to the 3M and 5M modulations.
These suggest possibilities of occurrence of new kinds of modulations leading to
precursor phenomena in these materials or that there may be more complicated
structures with co-existence of multiple modulated phases. Signatures of 7M modulated
phases have been observed experimentally \cite{mn2niga2,mn2niga3} in Mn$_{2}$NiGa, but
the occurrences of these were either dependent on the amount of stress in the system \cite{mn2niga3}
or on the sublattice occupancies \cite{mn2niga2}. Thus, no definite conclusion on the kind
of modulation in this system and the resulting pre-martensitic structures can be made
from the available experimental results. Detailed systematic calculations on the
non-cubic variants for these systems are to be carried out in order to settle the issue. However,
this is beyond the aim and scope of the present study.

Energetically lowest optical T$_{2g}$ branch is Raman active in nature with
[$\xi\bar{\xi}0$] polarization and the other optical branches are infrared
active with T$_{1}$u symmetry. It is known that phonon branches with
same symmetry would repel each other. Since, acoustic TA$_{2}$ branch also
has same state of polarization; it would be repelled by the T$_{2g}$
branches. In a previous theoretical study, Zayak {\it et al.} \cite{entel}
argued that due to this repulsion the TA$_{2}$ branch is pushed downward and
becomes unstable. To prove this, they compared the position of
T$_{2g}$ branches at $\Gamma$ point of some stable Heusler alloys at cubic
phase like Co$_{2}$MnGa and Co$_{2}$MnGe to unstable systems like
Ni$_{2}$Mn{\it X} ({\it X} =Ga, Ge, In, Al) and illustrated that energetically
lowered T$_{2g}$ branches in the unstable alloys compared to those alloys with
stable cubic phases, produce the necessary repulsive thrust to the lowest vibrational
branch. The results in FIG. 2 suggest the same explanation for the phonon instabilities
in Mn$_{2}$Ni{\it X}.
The repulsion due to the already low lying T$_{2g}$ modes at the $\Gamma$ point
for all four materials push the TA$_{2}$ frequencies down setting up the unstable modes.
In reference 30, the authors attributed the
occurrence of anomalous unstable modes in Ni$_{2}$MnGa to the inversion of modes of Ni
and Ga. They showed that the contributions to the T$_{2g}$ branches come from the
dynamics of Ni atoms and due to the inversion of optical modes, the Ni atoms vibrate
at lower frequencies making the frequencies of the T$_{2g}$ mode lower. The repulsion of
TA$_{2}$ modes by these T$_{2g}$ modes pull the frequencies of the former down making
them imaginary. For the materials investigated here, an analysis of the vibrational
amplitudes show that the T$_{2g}$ modes are dominated by the vibrations from Ni and
MnI atoms who occupy crystallographic equivalent sites, and in fact the same ones
as the two equivalent Ni atoms in Ni$_{2}$MnGa. Therefore, it would be interesting
to examine whether such an inversion of optical mode is also happening for these
materials. In the next subsection, we explore this by looking at the vibrational density
of states (VDOS).

\subsection{Vibrational density of states (VDOS)}
\begin{figure}[h]
\includegraphics[width=8.5cm]{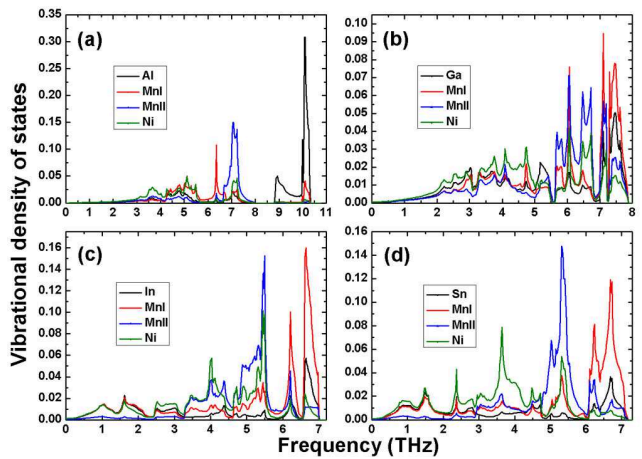}
\caption{(Color online) Atom projected vibrational
density of states (VDOS) showing contributions from
different constituent atoms for (a) Mn$_{2}$NiAl,
(b) Mn$_{2}$NiGa, (c) Mn$_{2}$NiIn and
(d) Mn$_{2}$NiSn over the frequency range.}
\end{figure}
In what follows, the atom projected VDOS for Mn$_{2}$NiAl, Mn$_{2}$NiGa,
Mn$_{2}$NiIn and Mn$_{2}$NiSn are presented in FIG. 3. It is observed that
the vibrational contributions from two Mn atoms occupy different frequency
regions in the VDOS plots. This occurs mainly because of the following reasons:
the two Mn atoms have different crystallographic symmetry; the atom
occupying (000) sub-lattice, labeled as MnI, have tetrahedral symmetry and
the other one at (0.25,0.25,0.25), labeled as MnII, sub-lattice have octahedral
symmetry; as a consequence of this their nearest neighbor environments are different leading to
different bond stiffness's (force constants) for the bonds connected to the Mn atoms. A comparison of
all the VDOSs show that the VDOSs of Mn$_{2}$NiIn and
Mn$_{2}$NiSn materials are quite similar and are very different from the VDOSs of
the other two materials in the series.
FIG. 3 suggests that for Mn$_{2}$NiIn and Mn$_{2}$NiSn,
vibrations of MnI atoms are prominent between 6 THz to 7 THz,
whereas contributions from MnII atoms are predominantly lie
between 4.5 THz to 6 THz. Due to
the slightly larger atomic mass than Mn atom, Ni vibrations occur mostly between
2.5 THz to 4.5 THz. As expected, the lower frequency regions are
dominated by In and Sn because they have larger atomic masses than Ni and Mn.
For Mn$_{2}$NiGa, vibrations in the range 7 THz to
8 THz are mainly dominated by MnI atom, while vibrations
from 5.5 THz to 7 THz have contributions from MnII atoms. A strong
peak originated from MnI vibrations coinciding with a peak originating from
vibrations of MnII atoms
is also observed at 6 THz.
In the frequency range 3 THz to 5 THz, vibrations of Ni atoms are predominant
and the lowermost part of the spectrum is dominated by the vibrations of the Ga atoms.
The features in the VDOS of Mn$_{2}$NiAl is different than the other three. The modes due
to the vibrations of Al atoms occur at around 10 THz due to extremely light mass of Al.
The Ni modes also occur at lower frequencies, similar to the cases of the other three.
The vibrations of MnI and MnII atoms dominate the middle of the spectrum with their
respective peaks at 6.25 THz and 7.3 THz. In case of Ni$_{2}$MnGa, Zayak {\it et al.} \cite{entel}
showed that the positions of Ga and Ni contributions to the VDOS were ``inverted", that is, the vibrations of
the lighter Ni atoms were at frequencies lower than those of heavier Ga atoms. They connected this anomalous
mode inversion to the instability of the TA$_{2}$ modes of Ni$_{2}$MnGa. In case of the systems studied
here, the overall features in the VDOSs of all
four materials suggest that there is no signature of inversion of Ni (MnI) modes with
those of the modes from the element {\it X}. Thus the occurrence of unstable TA$_{2}$ modes
cannot be associated to this.

\subsection{Inter-atomic force constants}
\begin{figure}[h]
\includegraphics[width=8.5cm]{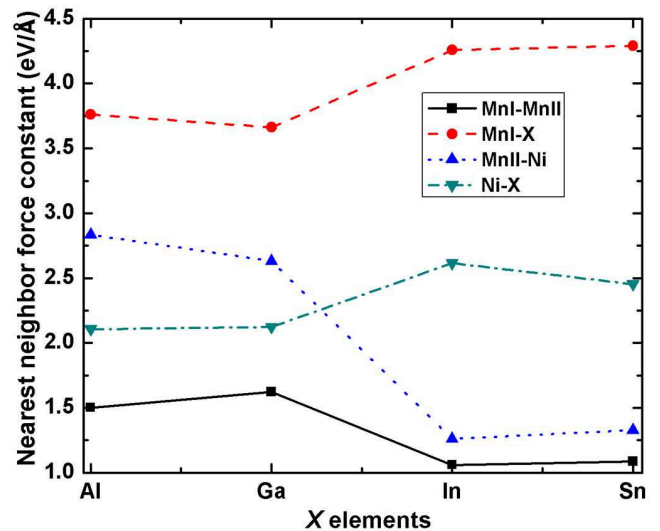}
\caption{(Color online) Longitudinal component of nearest neighbor
inter-atomic force constants between MnI, MnII, Ni and X atoms
of Mn$_{2}$Ni{\it X} materials.}
\end{figure}
In order to understand the features in the VDOS, we analyze the behavior of the real
space inter-atomic force constants.
In FIG. 4, we plot the longitudinal component of nearest
neighbor force constants of Mn$_{2}$Ni{\it X} systems. The transverse
components of force constants are not shown in the plot, since, their contributions
compared to the longitudinal ones are negligible.
The force constants between any pair of nearest neighbor
atoms are nearly equal for Mn$_{2}$NiAl with Mn$_{2}$NiGa. Same is true for Mn$_{2}$NiIn
with Mn$_{2}$NiSn. However, substantial changes in the force constants between any
pairs are observed as one moves from Mn$_{2}$NiGa
to Mn$_{2}$NiIn. Due to the increase in the inter-atomic distances, as a result of
expansion in their equilibrium lattice constants from 5.850 \AA \ to 6.162 \AA \,
the MnI-MnII and MnII-Ni longitudinal force constants become softer in Mn$_{2}$NiIn
and Mn$_{2}$NiSn in comparison to Mn$_{2}$NiGa. On the other hand,
the force constants related to {\it X} elements, i.e.,
MnI-{\it X} and Ni-{\it X} become harder in Mn$_{2}$NiIn and Mn$_{2}$NiSn
as compared to Mn$_{2}$NiGa and Mn$_{2}$NiAl.
This opposite behavior is observed since the sizes of the {\it X} elements for the former two
alloys are larger than those in the latter two, and thus are able to overcome the
expansion of the inter-atomic distances occurring in the former two as compared to the
latter two.
The nearest neighbor force
constants associated to MnII atom, the MnII-Ni and the MnII-MnI, become softer
as one moves from Mn$_{2}$NiGa to Mn$_{2}$NiIn and Mn$_{2}$NiSn. Therefore, vibration
frequencies corresponding to MnII atoms would be lower in the latter two materials, which
agree with the features in the VDOS. In Mn$_{2}$NiGa, vibrations of MnII extend from 5.5
THz to 7 THz, which in case of Mn$_{2}$NiIn and Mn$_{2}$NiSn shift to lower
frequencies, around 5.5 THz. The dynamical behavior of MnI and Ni atoms are
more complicated. For both of the atoms, two sets of inter-atomic force constants
behave opposite to one another. For Ni, the Ni-{\it X} nearest neighbor force constants
harden, as one goes from Ga to In and Sn. This should force Ni atoms to vibrate at higher
frequencies as one goes from Mn$_{2}$NiGa to Mn$_{2}$NiIn and Mn$_{2}$NiSn.
However, the vibrations of Ni atom remain more or less around the
same frequency for all the materials, since the previous effect is compensated by increasing
softening of the MnII-Ni bonds as one goes from Mn$_{2}$NiGa to Mn$_{2}$NiIn and Mn$_{2}$NiSn.
Similarly, hardening of
MnI-{\it X} force constant does not affect MnI vibrations, as this is compensated
by the softening
of MnI-MnII inter-atomic force constants.

\subsection{Fermi surfaces}
\begin{figure}[h]
\includegraphics[width=8cm]{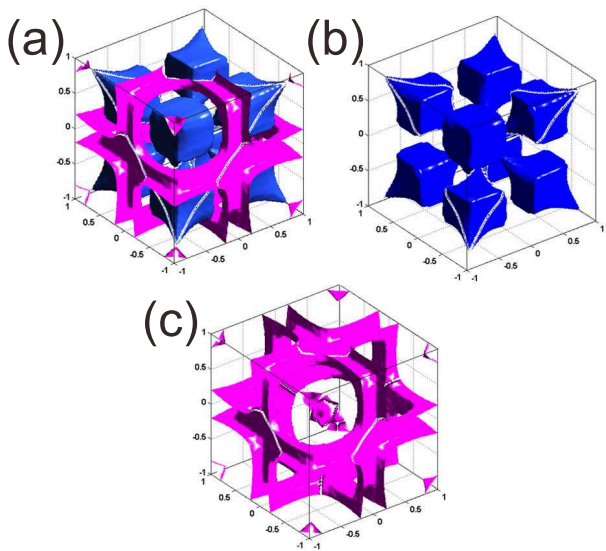}
\caption{(Color online) (a) Topology of 3D Fermi surfaces for Mn$_{2}$NiGa. The blue
and magenta surfaces represent 18$^{th}$ and 19$^{th}$ spin minority bands, respectively. (b)
and (c) illustrate those spin minority 18$^{th}$ and 19$^{th}$ bands separately.}
\end{figure}
\begin{figure}[h]
\includegraphics[width=8.5cm]{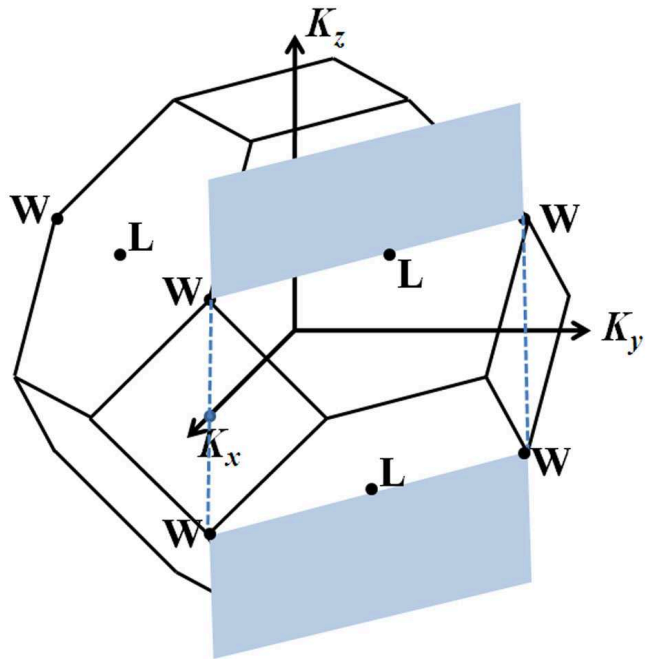}
\caption{(Color online) Illustration of the 110 cross section
(k$_{x}$+k$_{y}$=1) in fcc irreducible Brillouin zone (IBZ).}
\end{figure}

\begin{figure*}
\includegraphics[width=16cm]{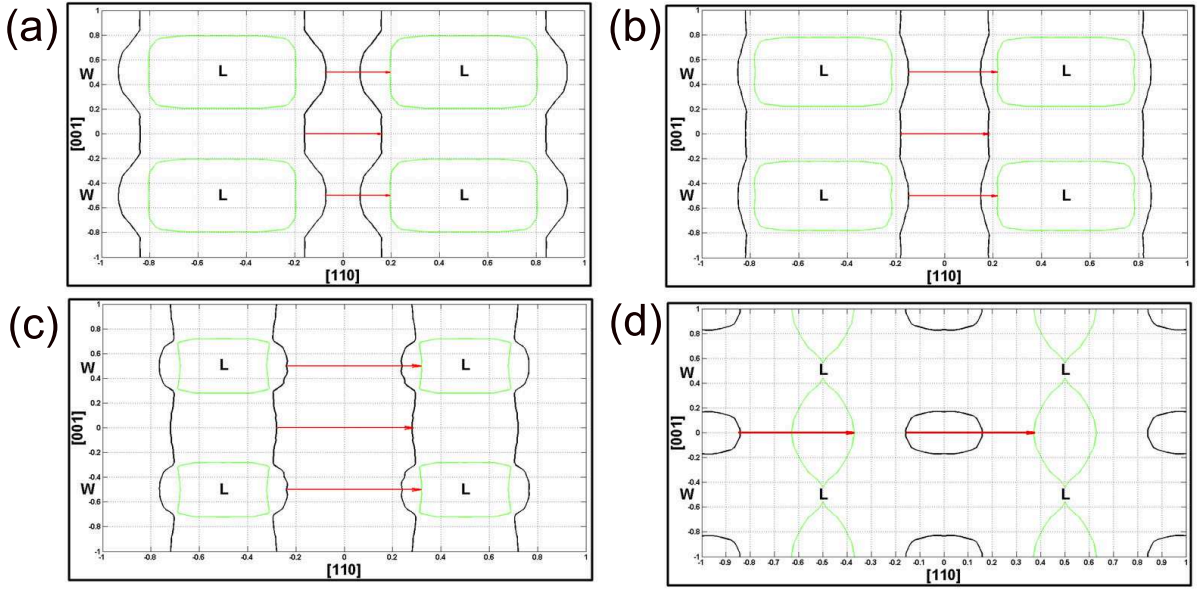}
\caption{(Color online) 2D cross section of the Fermi surfaces with the (110)
plane k$_{x}$+k$_{y}$=1 for (a) Mn$_{2}$NiAl, (b) Mn$_{2}$NiGa, (c) Mn$_{2}$NiIn
and (d) Mn$_{2}$NiSn. The green and black lines indicate spin minority bands. The red
arrows indicate nesting vectors {\bf q}=0.25(110) for Mn$_{2}$NiAl, {\bf q}=0.35(110)
for Mn$_{2}$NiGa and {\bf q}=0.50(110) for Mn$_{2}$NiIn and Mn$_{2}$NiSn.}
\end{figure*}
Previous first-principles studies in Ni$_{2}$Mn{\it X} relate the
martensitic instability of those materials with Fermi surface nesting \cite{entel,fs,fs1,fs2,fs3}.
The anomalies in the phonon branch mainly depend on the shape of the
Fermi surfaces and the electron-phonon matrix elements via the phonon
wave vector $\xi$ \cite{fs1,fs2}. This phenomenon occurs due to strong attraction between
two flat-parallel Fermi surfaces connected by a nesting vector \emph{\bf{q}}, at
the expense of atomic displacements and
at the wave vector where the maximum dip of the acoustic phonon branch is observed.
However, this cannot be generalized for all ternary alloys showing martensitic
instabilities. For Co$_{2}$NiGa, a newly found shape memory alloy,
Siewart {\it et al.} \cite{co2niga}
observed that softening in TA$_{2}$ phonon
branch was absent as a result of nonappearance of nesting features in the
Fermi surfaces of Co$_{2}$NiGa.
Here, we present Fermi surfaces corresponding to the spin-minority bands
only, since most prominent features are observed
in this spin channel as the systems undergo martensitic transitions \cite{pauljap}.
The three dimensional Fermi surfaces of Mn$_{2}$NiGa for
18$^{th}$ and 19$^{th}$ spin-minority bands are shown in FIG. 5.
The figure clearly exhibits flat portions of both the minority bands. However,
to examine the Fermi surfaces in details, to obtain clues about
the nesting between different parallel Fermi surfaces and hence, to relate this
novel feature to observed phonon anomaly, two dimensional (2D) projections
are necessary. In FIG. 7 we show the two-dimensional cross-sections of Fermi surfaces with the
(110) plane for the four systems (The relevant portion of the Irreducible Brillouin zone is 
shown in FIG. 6). The cross-sections for Mn$_{2}$NiGa, Mn$_{2}$NiAl and
Mn$_{2}$NiSn bear close resemblances while that of the Mn$_{2}$NiSn is somewhat different.
Inspite of this difference, the nesting vectors (indicated by red arrows in Fermi surfaces
plots) are consistent with the wave vectors at which the phonon anomalies are
observed in our phonon dispersion curves. Thus, we can conclusively
associate the occurrences of unstable modes in the Mn$_{2}$Ni{\it X} alloys with the
Fermi surface nesting. We refrain from further discussions on the differences in shapes of Fermi
surfaces between materials with the element $X$ belonging to different columns in the periodic
table because it is not necessary in the present discussion where the focus is on to establish
the nesting features in the Fermi surfaces and their relations to the martensitic instabilities
found in these systems.

In reference 36, Barman {\it et al.} also computed the Fermi surfaces of Mn$_{2}$NiGa. Surprisingly,
they observed Fermi surface nesting in the austenite phases along (100) and (010) directions
only, and not along (110) direction like we did. The $q$ value for one of the nesting vectors
found by them is quite close to ours (The $q$ value found by them is 0.31 which is very close to
our value, $q$=0.35) though. The nesting along (110) direction was observed by them in the martensitic phase with
the $q$ value 0.75. Though they
attributed this to the possible instabilities in the TA$_{2}$ phonon mode, it wasn't
substantiated by computations of the phonon spectra. Our results are qualitatively
different from theirs as we found nesting along (110) direction in the austenite phase of Mn$_{2}$NiGa. 
Moreover, our results are consistent as the Fermi surface
nesting along (110) could be related to the computed instabilities in the TA$_{2}$ phonon mode along
(110) with the nesting vector computed from the Fermi surfaces agreeing with the wave vector at which
the maximum of the instability occurs.

\subsection{Elastic constants}
\begin{table}[h]
    \caption{Calculated elastic constants and elastic anisotropy
    ratio for Mn$_{2}$Ni{\it X} materials. Experimental elastic constants
    are only available for Mn$_{2}$NiGa and shown in brackets.}
    \begin{center}
        \begin{tabular}{|c|c|c|c|c|c|} \hline
        Systems & c$^{\prime}$ & c$_{11}$ & c$_{12}$ & c$_{44}$ & A \\
                & (GPa)        & (GPa)    & (GPa)    & (GPa)    & (=c$_{44}/c^{\prime}$)\\ \hline
        Mn$_{2}$NiAl &-33.13 &100.35        &127.19         &131.66         &-3.97\\ \hline
        Mn$_{2}$NiGa &-13.42 &58.91         &125.17         &111.00         &-8.27\\
                     &       &(90.55)\cite{ec} &(128.00)\cite{ec} &(124.42)\cite{ec} &      \\ \hline
        Mn$_{2}$NiIn & 16.44 &118.64        & 85.76         & 41.47         &2.52\\ \hline
        Mn$_{2}$NiSn & 15.43 &146.05        &115.19         & 64.27         &4.17\\ \hline
        \end{tabular}
    \end{center}
\end{table}
The dynamical stability of crystalline phase implies that the strain
energy changes be positive definite against all possible small deformations.
This condition imposes restrictions on elastic constants. The stability
criteria for cubic crystals requires \cite{elconst}
\begin{eqnarray}
c_{44}> 0, c_{11}> |c_{12}|, c_{11}+2c_{12}> 0
\end{eqnarray}
Therefore to introspect the kinds of instabilities present in the
materials considered here and to validate our calculated phonon dispersion results,
we compute the elastic constants for all the four materials from the
initial slope ($\xi$$\rightarrow$ 0) of phonon dispersion plots along [$\xi\xi0$]
direction. The elastic constants c$_{44}$, c$^\prime$
(=$\frac{1}{2}(c_{11}-c_{12}$)) and c$_{L}$ (=$\frac{1}{2}(c_{11}+c_{12}+2c_{44}$))
are related to TA$_{1}$, TA$_{2}$ and LA acoustic modes \cite{elconst}. These elastic
constants are connected to ultrasound velocity via c$_{ij}$=$\rho\upsilon^{2}$ relation \cite{elconst}
where $\rho$ is the mass density. The three independent elastic constants of cubic crystal are
tabulated in TABLE I. Our computed c$_{12}$ and c$_{44}$ agree quite well
with the experimental results available only for Mn$_{2}$NiGa, whereas
in our calculation, c$_{11}$ is underestimated\cite{ec}. Overall the agreement with
experiment is good for Mn$_{2}$NiGa.
This, in effect, is an indirect indication to the accuracy of calculated
phonon spectra. The results show that the Equation (1) is satisfied by Mn$_{2}$NiIn and
Mn$_{2}$NiSn only. This indicates that Mn$_{2}$NiAl and Mn$_{2}$NiGa are unstable in the
cubic structure. We gain further insight into the nature of stabilities of these materials
by looking at the other two parameters listed in TABLE I, the shear constant and the elastic
anisotropy ratio. Since acoustic TA$_{2}$ branch is related to shear
constant (c$^{\prime}$), hence, negative c$^{\prime}$ for Mn$_{2}$NiAl and
Mn$_{2}$NiGa is an indication of pure elastic instability which stabilizes
though shear deformation across ($\xi\xi0$) planes in [$\xi\bar{\xi}$0] direction. The
same is not true for the other two materials. Although they satisfy Equation (1) and
have sizable c$^{\prime}$, their anisotropy ratios A are high enough to bring in a
martensitic transformation \cite{niti}.
The elastic anisotropy ratio A (=c$_{44}/c^{\prime}$) is an important
quantity to measure of stability of cubic structures under stress across ($\xi\xi0$)
planes \cite{zener}. Larger the value it acquires, more unstable the structure becomes.
For systems undergoing martensitic transformations, the value of A varies from 2 onward
\cite{niti,cunial,shapiro1,shapiro2,nimnsn-sb,acet}. In cases of Mn$_{2}$NiIn and Mn$_{2}$NiSn,
the values of A lie well within the limits observed in shape memory alloys. The origin of this
could be rather small value of the shear modulus c$^{\prime}$. Additionally, we find that
c$_{44}$ in cases of Mn$_{2}$NiIn and Mn$_{2}$NiSn are much softer than those for the other
two materials. The comparative softening in c$_{44}$ for Mn$_{2}$NiIn and Mn$_{2}$NiSn as
compared to Mn$_{2}$NiGa and Mn$_{2}$NiAl, indicate that the cubic Mn$_{2}$NiIn and Mn$_{2}$NiSn will transform
to different martensitic phases compared to the other two where the transformations would be driven by
softening in c$^{\prime}$ as has been observed in cases of other shape memory alloys \cite{niti}.
The results on elastic constants therefore corroborate the inferences drawn from the differences
in dispersion relations for the materials studied.

The vibrational and elastic properties discussed in this work show a clear trend. Mn$_{2}$NiGa
and Mn$_{2}$NiAl are quite similar in their behaviors; same goes for Mn$_{2}$NiIn and Mn$_{2}$NiSn.
The vibrational and elastic properties among these two groups are significantly different. The origin
of such differences can be traced back to the differences in their electronic structures \cite{pauljap}.
The signatures of mechanical instability were reflected in electronic structures of Mn$_{2}$NiGa and Mn$_{2}$NiAl,
where high densities of states, as compared to Mn$_{2}$NiIn and Mn$_{2}$NiSn, were found at the Fermi level.
The origin of this was larger hybridizations between the Mn and Ni atoms at the octahedral positions for the
former two systems. For the latter two systems, rather small densities of states at Fermi level, due to smaller
hybridizations between the magnetic atoms at octahedral positions, originating from larger distances between those
magnetic atoms (due to the atoms sitting in a larger lattice compared to the former two which happens as In and Sn have larger sizes
than Ga and Al), signified that it would take external influences to induce instabilities into these systems.

\section{Summary and Conclusions}
We have investigated the lattice dynamics of Mn$_{2}$Ni{\it X}
({\it X}= Al, Ga, In, Sn) MSMAs in their austenite phase using first-principles
based density functional theory calculations. The calculated phonon spectra show
anomalous behavior of the
acoustic TA$_{2}$ branch along [$\xi\xi0$] direction for
all the four materials indicating structural instability. Instabilities in the
said acoustic mode can be related to the repulsion by the optical T$_{2g}$ mode
having the same symmetry as the TA$_{2}$ mode. Unlike Ni$_{2}$MnGa, no inversion of
optical modes could be observed, thus ruling this out as one of the possible mechanisms
behind the anomalous features in phonon spectra. The features in the vibrational
densities of states can be explained from the qualitative variations of the interatomic
force constants across the materials. The calculated elastic constants corroborate the
structural instabilities inferred from phonon dispersion relations. Negative shear
constants for Mn$_{2}$NiAl and Mn$_{2}$NiGa indicate pure elastic instabilities in
these materials. Finally, the nesting features in the Fermi surfaces confirm that the
observed phonon anomalies are associated with them. The wave vectors at which the
maximum anomaly occur indicate the possibility of formation of pre-martensitic
modulated phases which are yet to be confirmed by experiments. The results also indicate
that these modulated pre-martensitic phases could be quite complex and further
investigations into this aspect is necessary.

\section{Acknowledgments}
Financial assistance from the Swedish Research Links (VR-SIDA) is acknowledged. The
Swedish National Computing facilities, computation facilities from C-DAC, Pune, India
and from Department of Physics, IIT Guwahati funded under the FIST programme of DST, India
are also acknowledged. SG and SP would like to acknowledge Dr. Munima B. Sahariah, IASST,
Guwahati, India for the help in plotting the Fermi surfaces.

\end{document}